\documentclass[aps,prl,superscriptaddress,amsmath,amssymb,reprint,longbibliography]{revtex4-2}

\usepackage{graphicx}% Include figure files
\usepackage{dcolumn}% Align table columns on decimal point
\usepackage{bm}% bold math
\usepackage{xcolor}

\usepackage{upgreek}

\usepackage[separate-uncertainty]{siunitx}
\usepackage[breaklinks=true]{hyperref}
\hypersetup{
	colorlinks=true,
	linkcolor=blue,
	citecolor=blue,
	urlcolor=blue	
}

\newcommand{\micron}{{\upmu\mathrm{m}}}
\newcommand{\Wcmsqd}{{\mathrm{W }/\mathrm{cm}^{2}}}
\usepackage[dvipsnames]{xcolor}
\usepackage{adjustbox}
\usepackage{graphicx}

\begin{document}

\title{
Self-organized positron reorienting and pinching mechanism for the experimental detection of the linear Breit-Wheeler process
%Self-organized linear Breit-Wheeler positron pinching mechanism in single-pulse laser-plasma interaction\\
%Generation of high linear Breit-Wheeler signal by self-organized positron pinching mechanism in single-pulse laser-plasma interaction}
}

\author{Yutong He}
\email{yutong.he@york.ac.uk}
\affiliation{York Plasma Institute, Department of Physics, University of York, York, YO10 5DD, United Kingdom}
\affiliation{GoLP/Instituto de Plasmas e Fusão Nuclear, Instituto Superior Técnico, Universidade de Lisboa, 1049-001 Lisboa, Portugal}

%\author{Matthew Idso}
%\affiliation{Department of Mechanical and Aerospace Engineering, University of California San Diego, La Jolla, CA 92093, USA}

%\author{George Herrera}
%\affiliation{Department of Mechanical and Aerospace Engineering, University of California San Diego, La Jolla, CA 92093, USA}
%\affiliation{General Atomics, San Diego, California 92186, USA}

\author{Alexey Arefiev}
\affiliation{Department of Mechanical and Aerospace Engineering, University of California San Diego, La Jolla, CA 92093, USA}

\author{Mario Manuel}
\affiliation{General Atomics, San Diego, California 92186, USA}

\author{Hui Chen}
\affiliation{Lawrence Livermore National Laboratory, Livermore, California 94550, USA}

%\author{L. O. Silva}
%\affiliation{GoLP/Instituto de Plasmas e Fusão Nuclear, Instituto Superior Técnico, Universidade de Lisboa, 1049-001 Lisboa, Portugal}

\author{Christopher Ridgers}
\email{christopher.ridgers@york.ac.uk}
\affiliation{York Plasma Institute, Department of Physics, University of York, York, YO10 5DD, United Kingdom}

\date{\today}

\begin{abstract}
The linear Breit-Wheeler (LBW) process ($\gamma+\gamma\rightarrow e^{-}+e^{+}$) is a fundamental prediction of quantum electrodynamics, but yet to be observed under laboratory conditions using real photons.
In recent years, a few experimental schemes utilizing high-intense ($\sim10^{22}$W/cm$^2$) laser-plasma interactions to observe the LBW process have been proposed. However, a high level of signal-to-noise-ratio are expected in these schemes, hindering the first-ever experimental detection of the LBW process by real photons. In this paper, we present a simple experimental setup which could enhance the expected positron signals by 2-3 orders of magnitude compared to previously proposed schemes, reaching the level of $10^{6}$~MeV$^{-1}$str$^{-1}$. Moreover, such high positron signal is achieve in the direction opposite to the laser propagation, where a significantly quieter background is expected compared to the previously focused direction of laser propagation. The key to achieve this result is a newly discovered self-organized positron reorienting and pinching mechanism, enabled by the in-situ strong plasma fields from the laser-plasma interaction.
%Recent studies proposed experimental schemes 
%
%Previous research [New J. Phys. 23, 115005 (2021), Phys. Rev. Lett. 131, 065102 (2023)] have shown that a pronounced amount of LBW pairs can be produced by irradiating an ultra-intense laser pulse into a foam target. Our numerical simulations show that the majority of the produced positrons in such a system can be trapped and pinched by the in-situ plasma fields, forming a backward-moving collimated positron beam reaching 106 str-1MeV-1 with an opening angle of around 30° in the near hundred MeV regime. Such trapping is due to the strong azimuthal plasma magnetic fields created by the laser-driven electron current, leading to the backward drifting of the LBW positrons. Consequently, a self-organized pinching mechanism is enabled by the gradual decrease of such magnetic fields experienced by the LBW positrons. Our results suggest a potentially feasible experimental regime for the first-ever detection of the LBW process under laboratory conditions using real photons, almost a century after this process was theoretically proposed.
%
\end{abstract}

\maketitle

\textbf{\textit{\large Introduction}}

Proposed almost a century ago in 1934, the linear Breit-Wheeler (LBW) process ($\gamma+\gamma\rightarrow{e^{-}}+{e^{+}}$) is one of the most basic predictions of quantum electrodynamics (QED)~\cite{breit.PhysRev.1934}.
This process is believed to play a key role in revealing some of the biggest mysteries in contemporary astrophysics, such as the fast radio bursts~\cite{he.arxiv.2025} and pulsar radio emission~\cite{Harding.rpp.2006}, and have other important applications in astrophysical scenarios such as the relativistic radiation mediated shocks~\cite{Beloborodov.apj.2017,levinson.PRE.2020}, the gamma ray bursts~\cite{levinson.PhysRep.2020}, or the black hole magnetosphere magnetosphere~\cite{beskin.az.1991,Hirotani.apj.1998} and black hole jets~\cite{Broderick.apj.2015}.
However, the LBW process has never been observed under laboratory conditions with real photons.
The lack of experimental detection leaves this rudimentary part of QED theory, one of the most successful and accurate physical theories developed, undertested. Moreover, it rules out the opportunity of studying these intriguing astrophysical puzzles in a laboratory.

The difficulty for experimentally realizing the LBW process comes from the requirement for the collision of photons that are both dense and energetic, which none of the traditional photon sources is capable to achieve.
%The high photon energy threshold ($\sim$~MeV) and the extremely small cross section ($\sim10^{-30}$~m$^{2}$) of the LBW process lead to the experimental requirement for the collision of photons that are both dense and energetic, which none of the traditional photon sources is capable to achieve.
On the other hand, the rapid development of high-power ($\sim$PW) and high-intensity ($\sim10^{22}$~W/cm$^{2}$) laser technology~\cite{Danson.hplse.2019,Lureau.hplse.2020,Yoon.optica.2021} opened the new avenue of experimenting the LBW process by the newly available high-energy photon sources produced by ultra-intense laser pulses.
Theoretical and computational efforts have been conducted over the past decade where various experimental schemes have been proposed~\cite{Pike.np.2014,Ribeyre.PRE.2016,wang.PhysRevApplied.2020,he.CommPhys.2021,golub.prd.2021}. These schemes require the use of multiple laser pulses, which is experimentally challenging.
Breakthrough was achieved in the recent few years where experimental schemes using only a single laser pulse were proposed~\cite{he.njp.2021,sugimoto.prl.2023,song.arxiv.2025}.

However, these currently proposed single-pulse schemes still suffer from limitation on signal to noise ratio, which severely hinder the first-ever detection of the LBW pair creation with real photons in history.
To produce the required dense and energetic photons for the LBW pair creation, the required laser intensity in these schemes is extremely high ($\sim10^{22}$~W/cm$^{2}$). A glaring background noise is thus expected. Despite such a power laser pulse is used, these schemes could only produce expected positron signals ($\sim10^{3-4}$~str$^{-1}$MeV$^{-1}$) which are low.
As a comparison for reference, 
in previous laser-plasma experiments of pair creation (via the Bethe-Heitler process), laser pulses of much lower intensity ($\sim10^{18-19}$~W/cm$^{2}$) were used~\cite{gahn.pop.2002,chen.pop.2009,williams.pop.2015,xu.pop.2016}, producing a much quieter background. Nevertheless, the positron detection threshold ($\sim10^{4-7}$~str$^{-1}$MeV$^{-1}$) in these experiments were still orders of magnitude higher than the expected LBW positron signals in the proposed schemes.
Therefore, major improvements that can significantly enhance the expected LBW positron signal, meanwhile systematically suppress the background noise has to be achieved.

In this paper, we propose a simple experimental scheme that could accomplish this goal. We show that during the interaction of a currently available laser pulse with a plastic foam target, a self-organized positron reorienting and pinching mechanism can be induced by the in-situ strong plasma fields.
The pinching effect enhances the positron signal by orders of magnitude, and remarkably, the positrons are redirected to the direction opposite to the laser propagation (referred as ``backward'' hereafter).
Compared to the direction of laser propagation (referred as ``forward'' hereafter), the backward direction typically has a much quieter background, which can significantly suppress the noise level. By 3D particle-in-cell (PIC) simulations, we show that an expected LBW positron signal of $\sim10^{6}$~str$^{-1}$MeV$^{-1}$ can be achieved in the backward direction. Such positron signal is around 2-3 orders of magnitude higher than in previously proposed schemes, which enters the detectable range of positron signals in those previous laser-plasma experiments~\cite{gahn.pop.2002,chen.pop.2009,williams.pop.2015,xu.pop.2016}.
%More crucially, the expected number of MeV-level photons and electrons co-propagating with these positrons could be reduced by approximately $\ytt{(xx)}$ orders of magnitude, leading to an total improvement on the positron signal over noise signal by approximately $\ytt{(xx)}$ orders of magnitude.

\textbf{\textit{\large Positron Reorienting and Pinching}}

\begin{figure}
\includegraphics[width=\linewidth]{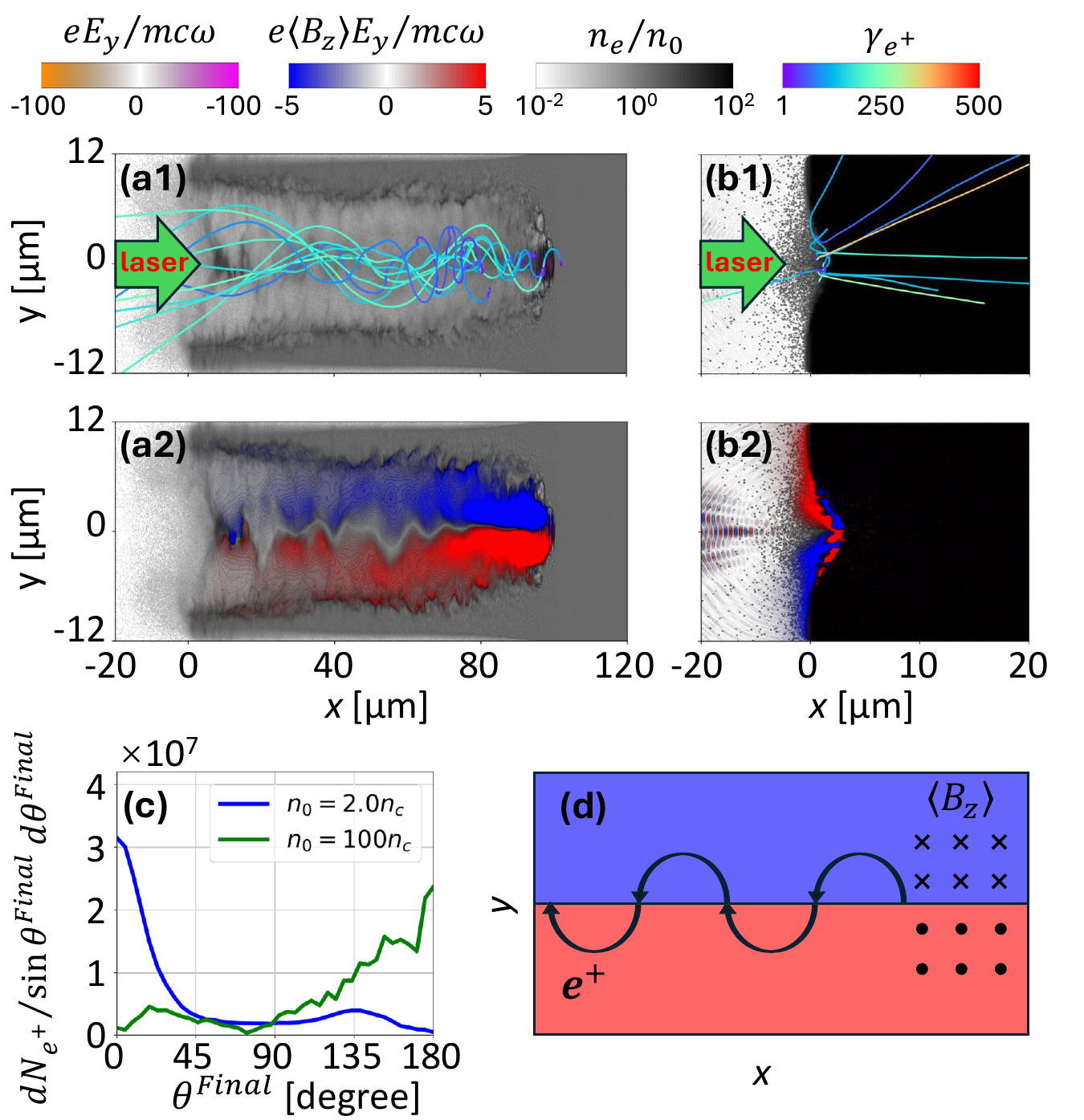}
\caption{
\textbf{(a1, b1)} Injection of a laser pulse into a uniform target with initial density $n_0=2n_c$ (a1) and $n_0=100n_c$, respectively. The color-coded curves shows the trajectories of 10 randomly selected positrons projected into the $(x-y)$ plane, with the color coding representing the positron Lorentz factor $\gamma$.
\textbf{(a2, b2)} $\langle B_z\rangle$, z-component of the magnetic fields in the $z=0$ plane averaged over 4 laser periods, normalized by $mc\omega/e$.
The gray scale in (a1, a2, b1, b2) shows the plasma density $n_e$ in the $z=0$ plane, normalized by $n_0$.
\textbf{(c)} Distribution $dN_{e^+}/\sin\theta^{Final}d\theta^{Final}$ of the positrons over their final polar angle $\theta^{Final}$ in systems with $n_0=2n_c$ and $n_0=100n_c$.
\textbf{(d)} Schematic illustration of the backward drifting of the produced positrons by the azimuthal magnetic field.
} \label{fig:1}
\end{figure}

To demonstrate the key idea of the mechanism, we first compare the angular distribution of the LBW positrons in two simple systems.
As illustrated in panel (a1,a2,b1,b2) of Fig.~\ref{fig:1}, 3D PIC simulations are performed where a laser pulse of wavelength $\lambda_0=800$~nm and peak intensity $I_0=3\times10^{22}$~W/cm$^2$ is injected into a uniform C-H target of initial density $n_0$ being $2n_c$ (panel (a1,a2)) and $100n_c$ (panel (b1,b2)). Here $n_c\equiv m\omega_0^2/4\pi e^2$ is the critical plasma density with $m$, $e$, and $\omega_0$ being the electron mass, elementary charge, and laser frequency $\omega_0\equiv2\pi/\lambda_0$, respectively. The laser intensity corresponds to a normalized laser amplitude of $a_0=eE_0/mcw_0=120$ with $E_0$ being the peak laser field and $c$ being the speed of light.
We define $x$ to be the direction of laser propagation. Details of the system and simulation setup is provided in Appendix B.
In these simulations, ultra-relativistic electrons are generated as the intense laser pulse interacts with the plasma. Dense photons in multi-MeV energy range are then emitted by these energetic electrons via the synchrotron emission. These photons subsequently can collide with each other and produce LBW pairs.

These two systems produce close pair yields ($7.2\times10^{6}$ in the $n_0=2n_c$ system, and $9.9\times10^{6}$ in the $n_0=100n_c$ system).
However, the angular distribution of the produced LBW positrons in these two systems are distinctively different.
To quantitatively illustrate such difference, we define the polar angle $\theta$ of a positron $\theta:=\arctan(-\sqrt{p_y^2+p_z^2}/p_x)$, with $p_{x,y,z}$ being the positron momentum in $x$, $y$, and $z$. $\theta=\pi$ corresponds to the perfect forward direction, and $\theta=0$ corresponds to the perfect backward direction.
Let $\theta^{Final}$ denote the final angle of a positron as it leaves the simulation box, and the final positron angular distribution $dN_{e^+}/\sin{\theta^{Final}}d\theta^{Final}$ is shown in panel (d) of Fig.~\ref{fig:1}.
This panel shows that most of the positrons in the $n_0=100n_c$ system propagate forward, as commonly seen in such laser-plasma systems as in Ref.~\cite{sugimoto.prl.2023,song.arxiv.2025}.
However, most positrons in the $n_0=2n_c$ system unexpectedly move backwards, and concentrates towards $\theta^{Final}=0$.

The key to understand the backward motion of the positrons in the $n_0=2n_c$ system is the configuration of the strong azimuthal plasma magnetic field. In the $n_0=2n_c$ system, the plasma is transparent to the laser pulse due to relativistic transparency. As the laser pulse propagates inside the plasma, a strong electron current is driven, co-propagating with the laser pulse. A strong azimuthal magnetic field is thus generated by this electron current. Such plasma B field is illustrated in Fig.~\ref{fig:1}(a2) which shows $\langle B_z\rangle$, the z-component of the magnetic field in the $z=0$ plane averaged over 4 laser periods. The produced positrons can then be confined by this B field, and reoriented to drift backward. Fig.~\ref{fig:1}(d) schematically illustrate such confining and reorienting effect, which is confirmed by the representative positron trajectories in the PIC simulation shown in Fig.~\ref{fig:1}(a1). In contrast, the opaque plasma in the $n_0=100n_c$ system prevents the formation of the azimuthal magnetic field, as shown in Fig.~\ref{fig:1}(b2). Thus most of the produced positrons simply move forward as shown in Fig.~\ref{fig:1}(b1).

The simple scheme shown in Fig.~\ref{fig:1}(d) can only explain the backward drifting of the positrons, and it is insufficient to explain the angular pinching towards $\theta^{Final}=0$ that we observe from Fig.~\ref{fig:1}(c).
This is because in this simple scheme, the polar angle $\theta$ of each positron would oscillate with the same amplitude during each oscillation of the positron trajectory. Thus no collective change of the positron polar angle is induced.

In the following, we present a simple analytical model to explain the angular pinching of the positrons.
The key is the structure of the azimuthal plasma B field $B_\phi$.
As the laser pulse propagates inside the plasma, the laser-expelled plasma channel expands transversely, forming the funnel-shape plasma channel shown in Fig.~\ref{fig:1}(a1,a2).
The current density inside the plasma channel thus has a non-zeros transverse component $j_r$, leading to a longitudinal gradient of the plasma B field: $\partial_x B_{\phi}\propto j_r\neq0$ (assuming $\partial_\phi=0$).
This can be seen in Fig.~\ref{fig:1}(a2) where $\langle B_\phi\rangle$ peaks around the laser front, and gradually decreases towards $-x$.
To illustrate the impact of $\partial_x B_\phi$ on positron dynamics, we define cylindrical coordinate $(r,\phi,x)$ with $r:=\sqrt{y^2+z^2}$ and $\phi$ being the radial and azimuthal components. Consider the funnel-shape current density $\Vec{j}=(j_r, j_\phi,j_x)$ such that $j_r>0$, $j_\phi=0$, and $j_x<0$, and assume symmetry in $\phi$: $\partial_\phi=0$. From basic electromagnetism, one can show such current configuration corresponds to a magnetic field $\Vec{B}$ which only has an azimuthal component: $\Vec{B}=(0,B_\phi,0)$ (see Appendix A for a detailed proof). The corresponding vector potential thus can be defined as $\Vec{A}=(A_x,0,0)$, with $A_x := -\int_0^r B_{\phi} dr'$. We choose this vector potential because it has the following important properties:
\begin{equation}
    A_x = 0 \;\;\; \text{and} \;\;\;  \partial_xA_x = 0 \;\;\;\;\;\; \text{at} \;\;\; r=0.\label{eq:1}
\end{equation}
%The orientation of the electron current determines the sign of $B_\phi$, which then ensures
%\begin{equation}
%    A_x >0 \;\;\;\;\; \forall r\neq0. \label{eq:2}
%\end{equation}
%Furthremore, from ...
From $\Vec{\nabla}\times(\Vec{\nabla}\times\Vec{A})=(4\pi/c)\Vec{j}$, we have $\partial_r(\partial_xA_x)=(4\pi/c)j_r>0$. Provided $\partial_xA_x=0$ at $r=0$ given by Eq.~(\ref{eq:1}), we have:
\begin{equation}
    \partial_x A_x >0 \;\;\;\;\;\; \forall r\neq0. \label{eq:3}
\end{equation}
We now consider the Hamiltonian $\mathcal{H}$ for a positron moving in this field: $\mathcal{H}=c[(\Vec{P}-e\Vec{A})^2+m^2c^2]^{1/2}$, with $\Vec{P}=\Vec{p}+e\Vec{A}$ being the canonical momentum. We first notice the two obvious integrals of motion: $d_tp_\phi=0$, and $d_t\gamma=d_t[p_x^2+p_r^2+p_\phi^2]^{1/2}/mc=0$, where $\gamma$ is the positron Lorentz factor. Thus $(p_x^2+p_r^2)$ is fixed. On the other hand, the definition of positron polar angle $\theta$ can be recast into $\theta:=\arctan[-\sqrt{p_r^2+p_\phi^2}/p_x]$. We thus have the important observation that $\theta$ is uniquely determined by $p_x$, and $\theta$ monotonically increases with the increase of $p_x$.
From Hamilton's equation $d_tP_x=-\partial_x\mathcal{H}$, we have:
\begin{equation}
    d_t(p_x+eA_x) = (e/\gamma m) \cdot p_x\cdot (\partial_x A_x) \label{eq:4}
\end{equation}
For a backward-drifting positron, the right hand side of Eq.~(\ref{eq:4}) is always non-positive, since $p_x<0$ and $\partial_xA_x \geq0$ from Eq.~(\ref{eq:1},\ref{eq:3}). So, $(p_x+eA_x)$ is monotonically decreasing in time.
When $r=0$, $A_x=0$ by Eq.~(\ref{eq:1}), thus $(p_x+eA_x)=p_x$. Therefore, each time a positron moves back to the central axis $r=0$, its $p_x$ will always be smaller than last time. Due to the relation between $p_x$ and $\theta$, this is true also for $\theta$: each time a positron moves back to the central axis $r=0$, its $\theta$ will always be smaller than last time.
On the other hand, during each oscillation, a positron will have a maximal polar angle at $r=0$. In another word, the oscillation amplitude of $\theta$ are determined by $\theta$ at $r=0$.
Thus the oscillation amplitude of $\theta$ will always decrease as a positron drifts backward. Collectively, the angular distribution of the positrons are pinched towards $\theta=0$ as they drift backward.
Finally, notice in the case $\partial_x=0$, as in the scheme shown by Fig.~\ref{fig:1}(d), the right hand side of Eq.~(\ref{eq:4}) is always zero. $p_x$ will then have the same value each time $r=0$, and the oscillation amplitude of $\theta$ is fixed. As expected, no pinching would occur in this case.

\begin{figure}
\includegraphics[width=\linewidth]{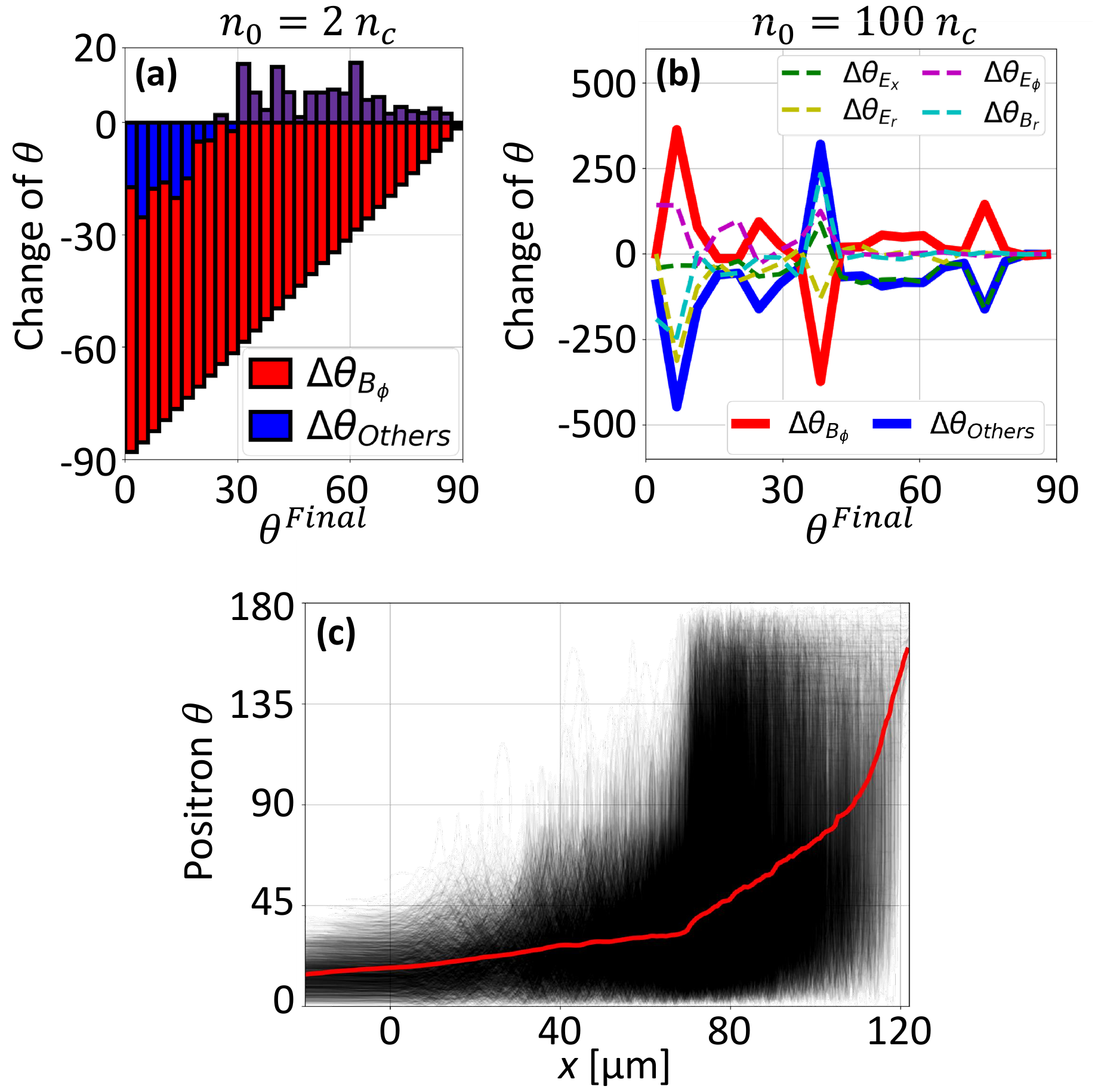}
\caption{
\textbf{(a,b)} Contribution to the change of $\theta$ by different field components in the $n_0=2n_c$ (a) and $n_0=100n_c$ (b) system. Here, $\Delta E_x$, $\Delta E_r$, $\Delta E_phi$, $\Delta B_r$, and $\Delta B_\phi$ are the contribution by $E_x$, $E_r$, $E_\phi$, $B_r$, and $B_\phi$, respectively, and $\Delta\theta_{Others}=\Delta E_x+\Delta E_r+\Delta E_\phi+\Delta B_r+\Delta B_\phi$. For each positron, the calculation starts from the point $\theta=\pi/2$.
\textbf{(c)} Trajectories in the $(x-\theta)$ plane of LBW positrons produced in the region $x>70\micron$ (gray curves). The red curve shows the averaged $\theta$ for each longitudinal coordinate $x$.
} \label{fig:2}
\end{figure}

To confirm that the plasma azimuthal magnetic field is the primary factor responsible for the angular pinching of the positrons, as in our analytical model. We performed a simple numerical diagnosis. We can decompose the change of $\theta$ by different field components:
\begin{eqnarray}
    \frac{d\theta}{dt} &=& - \frac{e}{\gamma mc} \frac{1}{|p_\perp|} \Big[p_\phi B_r-p_rB_\phi\Big] \nonumber \\
    && - \frac{e}{p^2}\Big[\frac{p_x}{|p_\perp|}(p_rE_r+p_\phi E_\phi) -|p_\perp|E_x\Big] \label{eq:5}
\end{eqnarray}
with $p_\perp$ being the positron momentum in the $(r-\phi)$ plane (equivalently, $(y-z)$ plane). A detailed derivation of Eq.~(\ref{eq:5}) is presented in Appendix C. We then numerically integrated the contribution to the change of $\theta$ by different field components in the PIC simulation.
Fig.~\ref{fig:2}(a,b) shows the contribution to the change of $\theta$ averaged over positrons with different values of $\theta^{Final}$. For each positron, the calculation starts from the time $\theta=\pi/2$ (i.e., perpendicular to the $x$ axis). Thus the presented change of $\theta$ is with respect to $\theta=\pi/2$. Panel (a) of Fig.~\ref{fig:2} confirms that in the system with $n_0=2n_c$, the azimuthal magnetic field $B_\phi$ is the primary contributor to the change of $\theta$, thus the factor responsible for the angular pinching of the positrons. In contrast, panel (b) of Fig.~\ref{fig:2} shows that the contribution to the change of $\theta$ by different field components in the $n_0=100n_c$ system is much more random, without a clear primary contributor.

Our analytical model also predicts that the pinching process of the positron occurs during their backward drifting. To confirm this prediction, in Fig.~\ref{fig:2}(c), we plot the trajectories in the $(x-\theta)$ plane of the positrons that are produced deep inside the plasma ($>70\micron$) in the $n_0=2n_c$ system. The gray shaded area filled by individual trajectories shrinks towards $-x$, confirming the decrease of the oscillation amplitudes of $\theta$ along $-x$. Moreover, the red curve in Fig.~\ref{fig:2}(c) shows the average positron $\theta$ at each longitudinal coordinate $x$, and it is confirmed in this panel that the averaged positron $\theta$ decreases along $-x$.

\textbf{\textit{\large Leveraging the positron signal}}

\begin{figure}
\includegraphics[width=\linewidth]{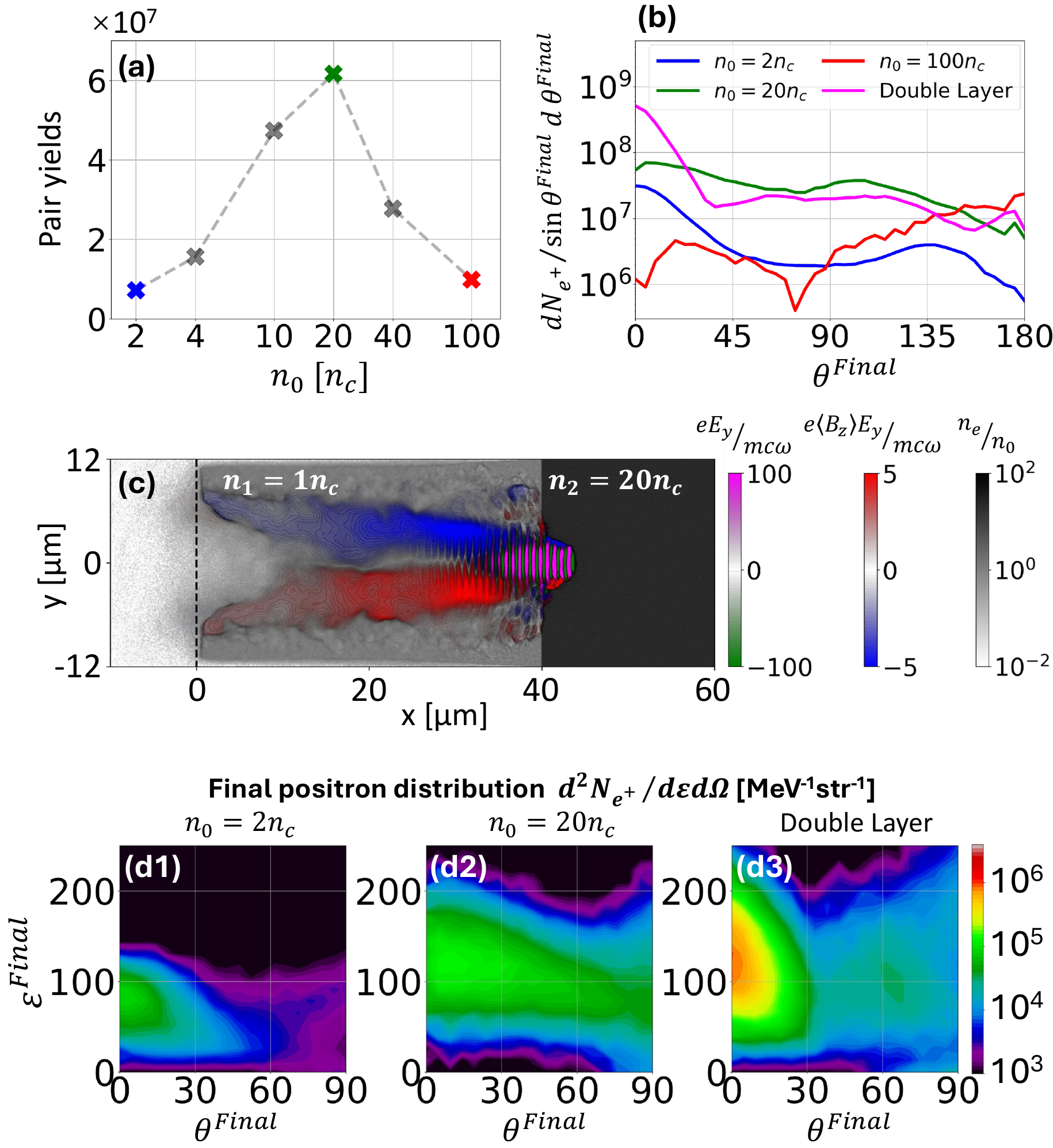}
\caption{
\textbf{(a)} Pair yields for systems with initial uniform plasma density being $2n_c$, $4n_c$, $10n_c$, $20n_c$, $40n_c$, and $100n_c$.
\textbf{(b)} Distribution $dN_{e^+}/\sin\theta^{Final}d\theta^{Final}$ of positron final angle for systems with $n_0=2n_c$, $20n_c$, $100n_c$, and the system with double layer target.
\textbf{(c)} Propagation of the laser pulse in the double layer target, with two layers of density being $n_1=1n_c$ and $n_2=20n_c$, respectively.
\textbf{(d1, d2, d3)} Final angular-energy distribution of the positrons $d^2N_{e^{+}}/d\varepsilon d\Omega$ [MeV$^{-1}$str$^{-1}$] for systems with $n_0$ being $2n_c$ and $100n_c$, and the double-layer system.
%
%dashed line.
} \label{fig:3}
\end{figure}

So far, we have identified a positron reorienting and pinching mechanism. To leverage the expected LBW positron signal, we performed a parameter scan over the initial uniform plasma density $n_0$ and compare the total pair yields. As illustrated in Fig.~\ref{fig:3}(a), by increasing $n_0$ from $2n_c$ to $20n_c$, the pair yield increases by approximately one order of magnitude (from $7.2\times10^6$ to $6.2\times10^{7}$). However, the positron signal in MeV$^{-1}$str$^{-1}$ cannot be increased by the same ratio. This is because as the plasma density increases, the plasma becomes more opaque. It is thus harder for the laser pulse to propagate inside the plasma and form the plasma magnetic field configuration needed to pinch the produced positrons. This is shown by the positron angular distribution in Fig.~\ref{fig:3}(b), where the green curve (representing the $n_0=20n_c$ system) only slightly surpass the blue curve (representing the $n_0=2n_c$ system) near $\theta^{Final}=0$. This panel also shows that for $n_0=100n_c$ (red curve), where the target is practically completely opaque, no backward drifting and pinching occurs, and the majority of the produced positrons have a forward final direction of motion.

Nevertheless, taking advantage of our newly identified pinching mechanism, we propose a simple solution to the inefficient pinching when the plasma density is relatively high. As illustrated in Fig.~\ref{fig:3}(c), the laser pulse is injected into a target consists of two layers, instead of a target with uniform density. The laser pulse first enters a layer of low density $n_1=1n_c$, where the plasma magnetic fields needed for efficient pinching is produced. Then, the laser pulse hits the second layer of higher density $n_2=20n_c$, where a pronounced amount of LBW pairs are produced. The total pair yield in this double layer system ($5.5\times10^{7}$) is similar to the pair yield in the $n_0=20n_c$ system ($6.2\times10^{7}$). However, as shown by the magenta curve in Fig.~\ref{fig:3}(b) near $\theta^{Final}$, the magnetic field configuration in the lower density layer retains the efficient positron pinching.
As a result, as shown in Fig.~\ref{fig:3}(d1, d2, d3), the positron signal in the double layer system can be further enhanced by approximately an order of magnitude compared to the ones in the uniform targets. In particular, a positron signal of approximately $10^6$~MeV$^{-1}$str$^{-1}$ can be achieved in the backward direction for an open angle of approximately $15^{\circ}$.

\textbf{\textit{\large Summary}}

To conclude, we discovered a self-organized positron reorienting and pinching mechanism during the interaction of an ultra-intense laser pulse with plastic foam target of near critical density. This mechanism can effectively reorient the produced LBW positrons to move towards the direction opposite to the laser propagation, where a significantly reduced background noise is expected. This mechanism can also angularly pinch the positrons towards the backward direction. In a simple setup we propose, the expected LBW positron signal can reach the level of $10^{6}$~MeV$^{-1}$str$^{-1}$, which is approximately 2-3 orders of magnitude higher than the positron signals in previous proposed laser-plasma schemes. Such positron signal already enters the positron detection range in previous laser-plasma experiments for pair creation, where the positrons were detected in the same direction of laser propagation.

The key component for the reorienting and pinching mechanism is the plasma magnetic fields created when an intense laser pulse propagates through a plasma with sufficiently low density ($\sim n_c$). In the setup we propose, this low-density plasma is realized by the placement of a layer of low density foam. In practice, this might be realized by other methods, such as the pre-plasma created by the pre-pulse.

%\clearpage\newpage

%\clearpage\newpage
%\appendix

\section{Appendix A}

In this appendix, we show that the quasi-static plasma magnetic field $\vec{B}$ only has an azimuthal component for the current density configuration considered in the paper. Specifically, we want to show that given any point $\Vec{r}=(r_r, r_\phi, r_x)\in\mathbb{R}^3$, one has $\Vec{B}(\Vec{r})=(0,B_{\phi}(\Vec{r}),0)$, provided the current density $\vec{j}$ does not have an azimuthal component: $\vec{j}=(j_r, 0, j_x)$, and symmetric in $\phi$: $\partial_\phi \vec{j} = (0,0,0)$.

\begin{figure}
\includegraphics[width=\linewidth]{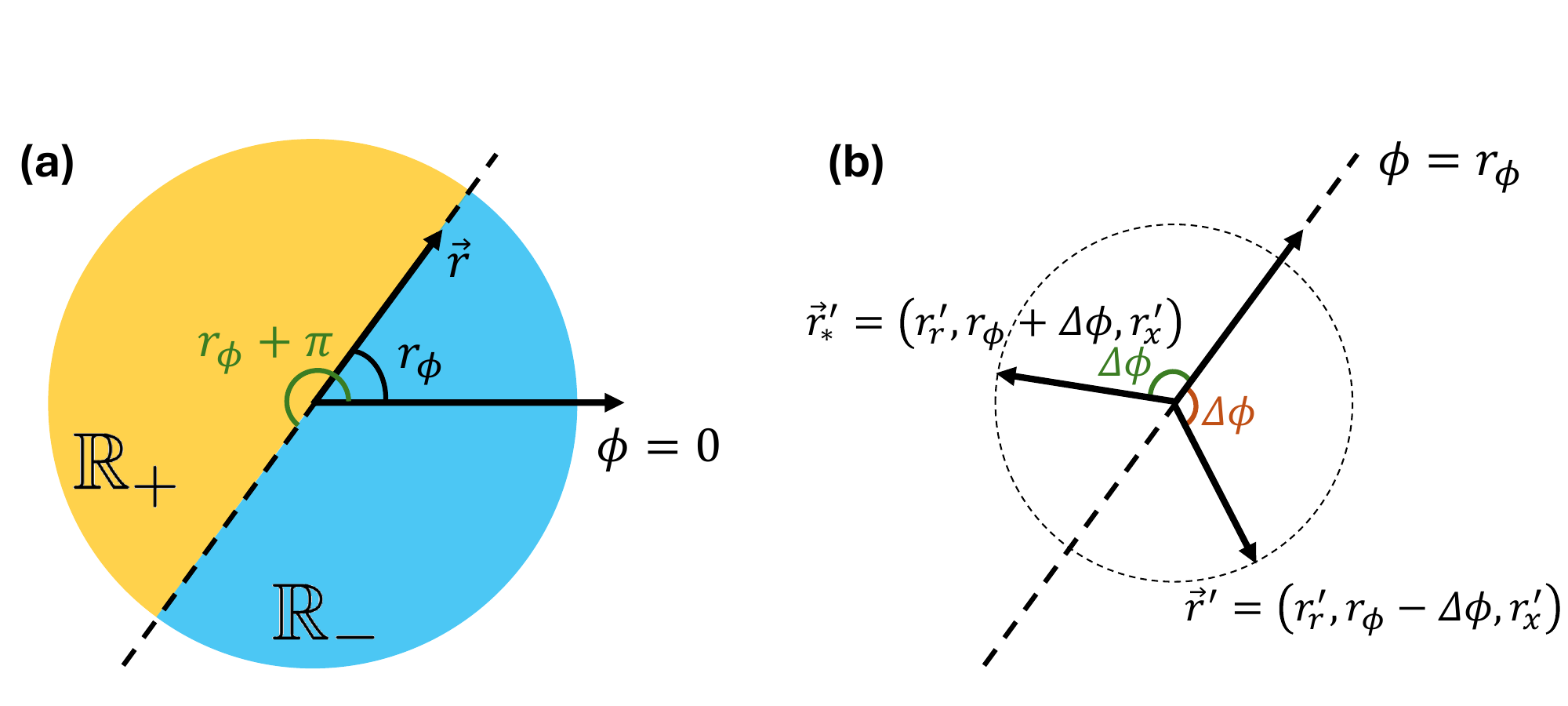}
\caption{\textbf{(a)} Schematic illustration of the definition of $\mathbb{R}_{-}$ (blue region) and $\mathbb{R}_{+}$ (yellow region).
\textbf{(b)} Schematic illustration of the definition of $\vec{r}'_*$ who is the reflection of $\vec{r}'$ with respect to $\phi=r_\phi$.
} \label{fig:s1}
\end{figure}

According to Biot-Savart law, we can express $\vec{B}(\vec{r})$ as the integration of the contribution from $\vec{j}$ across the space:
\begin{equation}
    \Vec{B}(\vec{r}) = \frac{1}{c} \int_{\mathbb{R}^3} \frac{\Vec{j}(\Vec{r}')\times(\Vec{r}-\Vec{r}')}{|\vec{r}-\vec{r}'|^3} d\vec{r}'.
\end{equation}
We first split the space $\mathbb{R}^3$ into two components $\mathbb{R}_{-}$ and $\mathbb{R}_{+}$ according to the azimuthal coordinate with respect to $r_\phi$:
\begin{eqnarray}
    \mathbb{R}_{-} &:=& \Bigl\{ (r,\phi,x)\in\mathbb{R}^3 \;\; \Big| \;\; \phi\in(r_\phi-\pi,r_\phi) \Bigr\} \\
    \mathbb{R}_{+} &:=& \Bigl\{ (r,\phi,x)\in\mathbb{R}^3 \;\; \Big| \;\; \phi\in(r_\phi, r_\phi+\pi) \Bigr\}
\end{eqnarray}
as illustrated in panel (a) of Fig.~(\ref{fig:s1}).
We can then split the integral into:
\begin{equation} \label{eq:s1.4}
    \Vec{B}(\vec{r}) = \frac{1}{c} \biggl(\int_{\mathbb{R}_{-}} +\int_{\mathbb{R}_{+}} \bigg) \frac{\Vec{j}(\Vec{r}')\times(\Vec{r}-\Vec{r}')}{|\vec{r}-\vec{r}'|^3} d\vec{r}'.
\end{equation}
For a point $\vec{r}'=(r'_r,r_\phi-\Delta\phi,r'_x)\in\mathbb{R}_{-}$, consider the corresponding point $\vec{r}'_* = (r'_r, r_\phi+\Delta\phi,r'_x)\in\mathbb{R}_{+}$ symmetric with respect to $\phi=r_\phi$, as illustrated in panel (b) of Fig.~(\ref{fig:s1}).
By definition, we have $(\vec{r}-\vec{r}')= (r_r-r_r',\; \Delta\phi,\; r_x-r_x')$, and $(\vec{r}-\vec{r}'_*)= (r_r-r_r',\; -\Delta\phi,\; r_x-r_x')$.
On the other hand, by symmetry, $|\vec{r}-\vec{r}'_*| = |\vec{r}-\vec{r}'|$, and $\vec{j}(\vec{r}'_*) = \vec{j}(\vec{r}')$. So:
\begin{eqnarray}
    && \int_{\mathbb{R}_{+}} \frac{1}{|\vec{r}-\vec{r}'_*|^3} \vec{j}(\vec{r}'_*) \times(\vec{r}-\vec{r}'_*) dv  = \int_{\mathbb{R}_{-}} \frac{1}{|\vec{r}-\vec{r}'|^3} \nonumber \\
    && \vec{j}(\vec{r}') \times (r_r-r_r',\; -\Delta\phi,\; r_x-r_x') d\vec{r}'.
\end{eqnarray}
Plug this into Eq.~(\ref{eq:s1.4}):
\begin{widetext}
\begin{eqnarray}
    \vec{B}(\vec{r}) && =\frac{1}{c} \int_{\mathbb{R}_{-}} \frac{1}{|\vec{r}-\vec{r}'|^3} \vec{j}(\vec{r}') \times \biggl[(r_r-r_r', -\Delta\phi, r_x-r_x') + (r_r-r_r',\; \Delta\phi,\; r_x-r_x') \bigg] d\vec{r}' \\
    && = \frac{1}{c} \int_{\mathbb{R}_{-}} \frac{1}{|\vec{r}-\vec{r}'|^3} \bigg(j_r,\; 0,\; j_x\bigg) \times \bigg(2(r_r-r_r'),\; 0,\; 2(r_x-r_x')\bigg) d\vec{r}'. \label{eq:s1.7}
\end{eqnarray}
\end{widetext}
The result of the cross product on the right hand side of Eq.~(\ref{eq:s1.7}) only has a $\phi$-component, thus $\vec{B}(\vec{r})$ also only has an azimuthal component.

\section{Appendix B} \label{sec:s2}

All of the PIC simulations discussed in this paper are performed in 3D-3V by the PIC code Epoch~\cite{Arber.ppcf.2015} with a Monte Carlo quantum synchrotron emission module~\cite{ridgers.jcp.2014} and a recently implemented LBW pair creation module. Parameters shared by all PIC simulations are listed in Table~\ref{table:1}. Other parameters that are specific to each particular simulation are listed in Table~\ref{table:2}. This paper contains PIC simulation results of 7 systems in total. These are systems of a uniform foam target with initial density $n_0[n_c]=2, 4, 10, 20, 40, 100$, and a double layer target. In Table~\ref{table:2}, we label the six uniform target systems by U002, U004, U010, U020, U040, U100, respectively, and the double layer target system by DL.

\renewcommand{\arraystretch}{1.15}
\begin{table}
\scriptsize

\centering
\begin{adjustbox}{width=\columnwidth}
\begin{tabular}{|c|c|}
 \hline
 \multicolumn{2}{|c|}{Laser parameters} \\
 \hline \hline
 Normalized peak amplitude & $a_0 = 120$ \\
 \hline
 Peak intensity& $I_0 = 3\times 10^{22}$~$\Wcmsqd$ \\
 \hline
 Wavelength & $\lambda_0 = 800$~nm \\
 \hline
 Focal plane of laser pulse & $x=0$ $\micron$ \\
 \hline
 Laser profile (longitudinal & Gaussian \\
 and transverse) & \\
 \hline
 Pulse duration (full width at & 25 fs\\
 half maximum for intensity) & \\
 \hline
 Focal spot size (full width at & 5.0 $\micron$\\
 half maximum for intensity) &  \\
 \hline
\end{tabular}
\end{adjustbox}

\bigskip

\scriptsize
\centering
\begin{adjustbox}{width=\columnwidth}
\begin{tabular}{ |c|c|  }
 \hline
 \multicolumn{2}{|c|}{Target parameters} \\
 \hline \hline
 Left boundary of target in $x$ & $x=0\micron$ \\
 \hline
 Target thickness (along $y$) & 24 $\micron$ \\
 \hline
 Target thickness (along $z$) & 24 $\micron$ \\
 \hline
 Target composition & $C^{+6}$, $H^{+}$, and electrons \\
 \hline
 Ion number density ratio & $n_{C^{+6}}: n_{H^{+}} = 1:1$ \\
 \hline
\end{tabular}
\end{adjustbox}

\bigskip

\scriptsize
\centering
\begin{adjustbox}{width=\columnwidth}
\begin{tabular}{ |c|c|  }
 \hline
 \multicolumn{2}{|c|}{Simulation parameters} \\
 \hline \hline
 Left boundary of  & $x=-20\micron$\\
 simulation box in $x$ & \\
 \hline 
 Simulation box in $y$ & $-12~\micron<y<12~\micron$\\
 \hline
 Simulation box in $z$ & $-12~\micron<z<12~\micron$\\
 \hline
 Spatial resolution & 20 cells per $\micron$ in $x$\\
 & 20 cells per $\micron$ in $y$\\
 & 20 cells per $\micron$ in $z$\\
 \hline
 Macro-particles per cell & 5 for electrons \\
 & 2 for carbon ions\\
 & 2 for hydrogen ions\\
 \hline
% Distance from particle probes to simulation boundaries & \\  
% Macro-photon cutoff energy & 10~keV \\
% \hline
% LBW up-sampling factor & 1000 \\
% \hline
\end{tabular}
\end{adjustbox}
\caption{Parameters shared by all 3D PIC simulations performed.}
\label{table:1}
\end{table}
\renewcommand{\arraystretch}{1}

%\begin{widetext}
\renewcommand{\arraystretch}{1.15}
\begin{table*}
\scriptsize
\centering
\begin{adjustbox}{width=\textwidth}
\begin{tabular}{|c|c|c|c|c|c|c|c|}
\hline
    System & U002 & U004 & U010 & U020 & U040 & U100 & \textcolor{white}{xxxx}DL\textcolor{white}{xxxx}  \\
\hline
    Initial plasma density & 2$n_c$& 4$n_c$& 10$n_c$& 20$n_c$& 40$n_c$& 100$n_c$ & 1$n_c$ for $0~\micron<x<40~\micron$ \\
        & & & & & & & 20$n_c$ for $40~\micron<x<70~\micron$ \\
\hline
    Target length (along x, & 120$\micron$ & 100$\micron$ & 55$\micron$ & 40$\micron$ & 40$\micron$ & 40$\micron$ & 70$\micron$ \\
        with left boundary placed at $x=0$) & & & & & & &  \\
\hline
    Right boundary of simulation box in $x$ & 122$\micron$ & 102$\micron$ & 57$\micron$ & 42$\micron$ & 42$\micron$ & 42$\micron$ & 72$\micron$ \\
\hline
\end{tabular}
\end{adjustbox}
\caption{Parameters specific to each 3D PIC simulations performed, with system names defined in Appendix B.}
\label{table:2}
\end{table*}
\renewcommand{\arraystretch}{1}
%\end{widetext}

\section{Appendix C}

In this appendix, we derive Eq.~(\ref{eq:5}), and present the method for calculating the change of positron polar angle by different field components in the PIC code.

Following the definition of the positron polar angle $\theta:=\arctan(|p_\perp|/(-p_x))$ with $p_\perp$ being the projection of $\Vec{p}$ onto the $y-z$ plane, the time derivative of $\theta$ is:
\begin{eqnarray}
    \frac{d\theta}{dt} &=& -\frac{d\arctan(|p_\perp|/p_x)}{dt} \\
    &=& -\frac{1}{p^2}\cdot\big[p_x(d_t|p_\perp|) - |p_\perp|(d_tp_x) \big].\label{eq:s3-1}
\end{eqnarray}
By definition, $|p_\perp|=(p_y^2+p_z^2)^{1/2}$, so:
\begin{equation}
    \frac{d|p_\perp|}{dt} = \frac{1}{|p_\perp|} \cdot \big[p_y(d_tp_y)+p_z(d_tp_z)\big].
\end{equation}
Plug this into Eq.~(\ref{eq:s3-1}):
\begin{eqnarray}
    && \frac{d\theta}{dt} = -\frac{1}{p^2}\cdot \nonumber\\
    && \bigg[\frac{p_x}{|p_\perp|}[p_y(d_tp_y)+p_z(d_tp_z)] - |p_\perp|(d_tp_x) \bigg]. \label{eq:s3-2}
\end{eqnarray}
On the other hand, from positron equation of motion, we can express $d_tp_x$, $d_tp_y$, and $d_tp_z$ in terms of the electromagnetic field components:
\begin{eqnarray}
    d_tp_x &=& eE_x +\frac{e}{\gamma mc} (p_yB_z-p_zB_y) \\
    d_tp_y &=& eE_y +\frac{e}{\gamma mc} (p_zB_x-p_xB_z) \\
    d_tp_z &=& eE_z +\frac{e}{\gamma mc} (p_xB_y-p_yB_x).
\end{eqnarray}
Plug these into Eq.~(\ref{eq:s3-2}), we arrive at Eq.~(\ref{eq:5}):
\begin{eqnarray}
    \frac{d\theta}{dt} &=& - \frac{e}{\gamma mc} \frac{1}{p^2} (\frac{p_x^2}{|p_\perp|}+|p_\perp|) \Big[p_zB_y-p_yB_z\Big] \nonumber \\ 
    && - \frac{e}{p^2}\Big[\frac{p_x}{|p_\perp|}(p_yE_y+p_zE_z) -|p_\perp|E_x\Big] \\
    &=& - \frac{e}{\gamma mc} \frac{1}{|p_\perp|} \Big[p_\phi B_r-p_rB_\phi\Big] \nonumber \\
    && - \frac{e}{p^2}\Big[\frac{p_x}{|p_\perp|}(p_rE_r+p_\phi E_\phi) -|p_\perp|E_x\Big] \label{eq:s3-3}
\end{eqnarray}

In the PIC code Epoch, at each time step and for each macro-positron, we can numerically integrate Eq.~(\ref{eq:s3-3}) to obtain the change of polar angle by each field component $\delta \theta_i$, using physical quantities (namely, momentum and Lorentz factor of the macro-positron, and fields experienced by the macro-positron) before the particle push. Here the index $i$ runs over each field component $E_x$, $E_r$, $E_\phi$, $B_r$, and $B_\phi$. Their sum $\delta\theta:=\sum_i \delta\theta_i$ thus represent the change of positron polar angle as predicted by Eq.~(\ref{eq:s3-3}). However, the PIC code Epoch uses the famous Boris pusher method~\cite{Arber.ppcf.2015,Boris.1970} to calculate the momentum update of charged particles by background electromagnetic fields. The change of positron polar angle $\widetilde{\delta\theta}$ by this method during the time step (i.e., the real change of $\theta$ in the PIC simulation) does not necessarily equal to $\delta\theta$. Furthermore, due to the specific form of this method, $\widetilde{\delta\theta}$ cannot be expressed as a linear combination of different field components as in Eq.~(\ref{eq:s3-3}).

We therefore make the correction by defining $\Delta\theta_i:=\delta\theta_i (\widetilde{\delta\theta}/\delta\theta)$, where we record $\Delta\theta_i$ in the code as the change of positron $\theta$ at this time step by each field component. Here, $\widetilde{\delta\theta}$ is obtained by comparing the positron polar angle (directed calculated from positron momentum as $\arctan(|p_\perp|/(-p_x))$) before and after the particle push. This correction ensures the consistency between Eq.~(\ref{eq:s3-3}) and the real change of $\theta$ in the simulation: $\sum_i \Delta\theta_i = \sum_i \delta\theta_i(\widetilde{\delta\theta}/\delta\theta) = \widetilde{\delta\theta}$, meanwhile retaining the ratio among $\delta\theta_i$, i.e., ensuring their relative contribution to the change of $\theta$ to be consistent with Eq.~(\ref{eq:s3-3}).

%\clearpage\newpage

\bibliography{apssamp}% Produces the bibliography via BibTeX.

%\section{End Matter}

\end{document}